\documentclass[12pt,english]{article}

\usepackage{CJKutf8} 
\usepackage[T1]{fontenc}
\usepackage{geometry}
\usepackage{stackengine}
\usepackage{siunitx}  
\usepackage{color}
\usepackage{pdfpages}
\usepackage{amsmath}
\usepackage{amsthm}
\usepackage{graphicx}
\usepackage{setspace}
\usepackage{float}
\usepackage{array}
\usepackage{multirow}
\usepackage{tabularx}
\usepackage{longtable}
\usepackage{tabu}
\usepackage{threeparttable}
\usepackage{adjustbox}
\usepackage{hyperref}
\usepackage{subcaption}
\usepackage{xurl}
\usepackage{comment}
\usepackage{tikz}
\usetikzlibrary{trees}
\usepackage[authoryear]{natbib}
\usepackage{hyperref}
\hypersetup{
    colorlinks=true,
    linkcolor=blue,
    filecolor=magenta,
    urlcolor=blue,
    citecolor=blue,
}
\newcolumntype{C}[1]{>{\centering\arraybackslash}p{#1}}
\newcolumntype{L}[1]{>{\raggedright\arraybackslash}p{#1}}
\newcolumntype{M}[1]{>{\centering\arraybackslash}m{#1}}

\makeatletter

\usepackage[section]{placeins}
\usepackage[english]{babel}
\usepackage{booktabs}
\usepackage{pdflscape}
\usepackage{afterpage}

\newenvironment{proof}[1][Proof]{\noindent\textbf{#1.} }{\ \rule{0.5em}{0.5em}}

\title{Will Compute Bottlenecks Prevent an Intelligence Explosion?}
\author{Parker Whitfill\thanks{MIT, whitfill@mit.edu}, Cheryl Wu\thanks{Yale University, cheryl.wu@yale.edu}}

\date{August 16, 2025}

\begin{document}
\maketitle

\begin{abstract}
\begin{singlespace}

The possibility of a rapid, "software-only" intelligence explosion brought on by AI's recursive self-improvement (RSI) is a subject of intense debate within the AI community. This paper presents an economic model and an empirical estimation of the elasticity of substitution between research compute and cognitive labor at frontier AI firms to shed light on the possibility. We construct a novel panel dataset for four leading AI labs (OpenAI, DeepMind, Anthropic, and DeepSeek) from 2014 to 2024 and fit the data to two alternative Constant Elasticity of Substitution (CES) production function models. Our two specifications yield divergent results: a baseline model estimates that compute and labor are substitutes, whereas a 'frontier experiments' model, which accounts for the scale of state-of-the-art models, estimates that they are complements. We conclude by discussing the limitations of our analysis and the implications for forecasting AI progress. 
\end{singlespace}
\end{abstract}

\setcounter{page}{1}
\begingroup
\renewcommand\thefootnote{}\footnote{We thank the Effective Altruism Forum's comments on the post version of this article. We would like to especially thank Basil Halperin and Philip Trammell for valuable feedback.}
\addtocounter{footnote}{-1}
\endgroup

\newpage
\section{Intro}
There has recently been significant advancements in the capabilities of Artificial Intelligence (AI) in domains such as coding and mathematics. Given that these skills are fundamental to AI research and development, this progress has raised the prospect of leveraging AI to accelerate AI research itself, a process termed recursive self-improvement (RSI). For example, Google Deepmind's AlphaEvolve is an LLM-based AI agent that discovered algorithmic advances that reduced LLM training time by $1\%$ \citep{DeepMind2025}.

Consequently, many industry insiders have argued that we are on the cusp of an intelligence explosion via recursive self-improvement - where AI perform AI research to train smarter models who do even more research to train even smarter models and so on. For example, the CEO of Meta described the prospect of an intelligence explosion as "compelling" \citep{Patel2025}. Similarly, the CEO of Anthropic noted that "because AI systems can eventually help make even smarter AI systems, a temporary lead could be parlayed into a durable advantage" \citep{Amodei2025}. Influential essays, such as `Situational Awareness' \citep{Aschenbrenner2024} and `AI-2027' \citep{Kokotajlo2025}, authored by former OpenAI researchers, project the emergence of super-intelligence through recursive self-improvement by the end of this decade and 2027, respectively. In academic economics, \cite{Halperin} investigates whether an intelligence explosion would also cause an explosion in economic output. 

Could AI's recursive improvement lead to a software-only intelligence explosion? Regarding this question, there have been hot debates about whether compute could bottleneck a software-only intelligence explosion \footnote{See Davidson's post for a summary: https://www.lesswrong.com/posts/XDF6ovePBJf6hsxGj/will-compute-bottlenecks-prevent-a-software-intelligence-1}. The rough idea is that AI research requires two inputs: cognitive labor and research compute. If these two inputs are gross complements, then even if there is recursive self-improvement in the amount of cognitive labor directed towards AI research, this process will fizzle as you get bottlenecked by the amount of research compute.

This compute bottleneck objection to a software-only intelligence explosion crucially relies on the assumption that compute and cognitive labor are gross complements. However, this fact is not at all obvious. On the one hand, compute and cognitive labor could be gross substitutes because more labor can substitute for a higher quantity of experiments via more careful experimental design or selection of experiments. For example, they can run generate better ideas and run small-scale experiments to optimize the experiments. On the other hand, skeptics of AI's recursive self-improvement believe that compute and cognitive labor are gross complements because eventually, ideas need to be tested out in compute-intensive, experimental verification. 

This paper investigates the conditions under which Recursive Self-Improvement in AI will trigger an intelligence explosion. We show  the extent to which `research compute' can substitute for the cognitive labor inherent in AI research. We explore the implications of this elasticity of substitution and the parameter governing the 'fishing out' effect for the feasibility of an intelligence explosion.

Theoretically, we demonstrate that two conditions are necessary for an 'intelligence explosion.' First, compute and labor must exhibit elastic substitution within the AI research production function. Second, the rate of AI self-improvement must outpace the increasing difficulty of discovering new ideas. We estimate $\sigma$ and find that the substitution between compute and labor in research is elastic in the baseline model but is inelastic in the frontier experiments model.

The rest of the paper proceeds as follows: section \ref{model} introduces the model, section \ref{data} introduces the data used in calibration, section \ref{results} shows the result, and section \ref{conclusion} concludes.

\section{Model}\label{model}
\subsection{Baseline CES in Compute}

We set up a theoretical model of researching better algorithms. 

Let $i$ denote an AI research firm and $t$ denote a time. Let $A_{it}$ denote the quality of the algorithms and $K_{it,\text{inf}}$ denote the amount of inference compute used by research firm $i$ at time $t$. We let $A_{it}K_{it,\text{inf}}$ denote effective compute \citep{ho2024algorithmic} for inference.

Algorithm quality improves according to the following equation \footnote{In reality, there is probably some algorithmic quality depreciation as firms scale up training compute (e.g., algorithms that are good for GPT-2 might be bad for GPT-4). We could accommodate this intuition by adding a term capturing depreciation.}:
\[
\dot{A}_{it} = \theta A_{it}^{\phi} F(K_{it,\text{res}}, L_{it})^{\lambda}.
\]

$\theta$ is the productivity scaling factor. $A_{it}^{\phi}$ denotes whether ideas, meaning proportional algorithmic improvements, get easier ($\phi > 1$) or harder to find ($\phi < 1$) as algorithmic quality increases, indexed by $A_{it}$. $F(K_{it,\text{res}}, L_{it})$ maps research compute $K_{it,\text{res}}$ and cognitive labor $L_{it}$ to a value representing effective research effort. $\lambda \le 1$ denotes a potential parallelization penalty.

We will assume $F(\cdot)$ is a constant-returns-to-scale production function that exhibits constant elasticity of substitution, i.e.,
\[
F(K_{it,\text{res}}, L_{it}) =
\begin{cases}
\left(\gamma K_{it,\text{res}}^{\frac{\sigma-1}{\sigma}} + (1-\gamma)L_{it}^{\frac{\sigma-1}{\sigma}}\right)^{\frac{\sigma}{\sigma-1}}, & \sigma \in (0,\infty),\ \sigma \neq 1, \\\\
K_{it,\text{res}}^{\gamma} L_{it}^{1-\gamma}, & \sigma = 1,
\end{cases}
\]
where $\sigma$ is the elasticity of substitution between research compute and cognitive labor. $\sigma > 1$ denotes the case where compute and cognitive labor are gross substitutes, $\sigma < 1$ where they are gross complements, and $\sigma = 1$ denotes the intermediate, Cobb-Douglas case.

Suppose that at time $t_0$, an AI is invented that perfectly substitutes for human AI researchers. Further, suppose it costs $c$ compute to run that system. Then $\dfrac{A_{it}K_{it,\text{inf}}}{c}$ denotes the number of copies that can be run.

We are interested in whether an intelligence explosion occurs quickly after the invention of this AI. We define an intelligence explosion as explosive growth in the quality of algorithms, $A_{it}$. Explosive growth of $A_{it}$ implies at least explosive growth in the quantity of AIs \footnote{If we additionally suppose that the intelligence of AI is an increasing, unbounded function of effective training compute, then explosive growth of $A_{it}$ would also imply explosive growth in AI intelligence.}.

Since we are interested in what happens in the short-run, we will assume all variables except algorithmic quality remain fixed. That is, we study if a software-only intelligence explosion occurs.

By assumption, the AI can perfectly substitute for human AI researchers at $t_0$. Therefore, effective labor dedicated to AI research becomes
\[
L_{it} = H_{it} + \frac{A_{it}K_{it,\text{inf}}}{c},
\]
where $H_{it}$ denotes human labor\footnote{As written, all inference compute is dedicated to AI research. We can easily weaken this assumption by having $c$ represent compute cost divided by the share of inference compute devoted to research.}. Plugging this effective labor equation into the equation that defines changes in algorithm quality over time:
\[
\dot{A}_{it} = \theta A_{it}^{\phi} F\!\left(K_{it,\text{res}},\, H_{it} + \frac{A_{it}K_{it,\text{inf}}}{c}\right)^{\lambda}.
\]

\subsection{Conditions for a Software-Only Intelligence Explosion}

The following are the necessary and sufficient conditions for explosive growth in $A_{it}$:
\[
\begin{cases}
\phi > 1, &\text{if } \sigma < 1,\\
\phi + (1-\gamma)\lambda > 1, & \text{if } \sigma = 1,\\
\phi + \lambda > 1, &\text{if } \sigma > 1.
\end{cases}
\]

To see why, let us go over the cases. If $\sigma < 1$, then the effective research effort term in our differential equation for $A_{it}$ is bounded. Intuitively, compute bottlenecks progress in effective research input. Therefore, the rate of growth of $A_{it}$ grows unboundedly if and only if the $A_{it}^{\phi}$ term grows over time, i.e., $\phi > 1$.

If $\sigma = 1$, then asymptotically we have
\[
\dot{A}_{it} = \theta A_{it}^{\phi} K_{it,\text{res}}^{\gamma\lambda} A_{it}^{(1-\gamma)\lambda}.
\]
We get hyperbolic growth if and only if $\phi + (1-\gamma)\lambda > 1$.

If $\sigma > 1$, then we are in the same case as $\sigma = 1$, except compute and cognitive labor are even more substitutable, so we drop the $(1-\gamma)$ term.

The $\sigma > 1$ condition is exactly what \citet{erdil2024estimating} consider when they analyze whether the returns to research are high enough for a singularity\footnote{There is a small notation difference as they denote the explosion condition as a fraction (e.g., $\lambda/(1-\phi) > 1$), while we express it as a sum.}. They find it possible that $\phi + \lambda > 1$, although the evidence is imperfect and mixed across various contexts. Therefore, if $\sigma > 1$, then a software-only intelligence explosion looks at least possible.

However, if $\sigma < 1$, then a software-only intelligence explosion occurs only if $\phi > 1$. But if this condition held, we could get an intelligence explosion with constant, human-only research input. While not impossible, we find this condition fairly implausible.

Therefore, $\sigma$ crucially affects the plausibility of a software-only intelligence explosion. If $\sigma > 1$ then it is plausible, but if $\sigma < 1$ it is not.

\subsection{Deriving the Estimation Equation}

We will estimate $\sigma$ by looking at how AI firms allocate research compute and human labor from 2014 to 2024.

Of course, throughout this time period, the AI firms have been doing more than allocating merely research compute and human labor. Their activities included training AIs and serving AIs, in addition to the research-focused allocation of compute and human labor.

Formally, they have been choosing a schedule of training compute $K_{it,\text{tra}}$, inference compute $K_{it,\text{inf}}$, research compute $K_{it,\text{res}}$ and human labor $H_{it}$. However, we can split the firm's optimization problem into two parts:
\begin{enumerate}
\item Dynamic Optimization: choosing $K_{it,\text{tra}}, K_{it,\text{inf}}, \bar{F}_{it}$ where $\bar{F}_{it} \equiv F(K_{it,\text{res}}, H_{it})$
\item Static Optimization: choosing $K_{it,\text{res}}, H_{it}$ to minimize costs such that $F(K_{it,\text{res}}, H_{it}) = \bar{F}_{it}$.
\end{enumerate}

In this split, we have assumed that $L_{it} = H_{it}$, i.e., that AIs did not contribute to cognitive labor before 2025\footnote{We think this assumption is very defensible for 2023 and prior. 2024 is borderline, so we try excluding it in our robustness test.}.

We will estimate $\sigma$ using the static optimization problem. Let $r_{it}, w_{it}$ denote the cost of research compute and human labor respectively. Then the static optimization problem becomes
\[
\min_{H_{it}, K_{it,\text{res}}} \; w_{it}H_{it} + r_{it}K_{it,\text{res}} \quad \text{such that } F(K_{it,\text{res}}, H_{it}) = \bar{F}_{it}.
\]

By taking the first-order conditions with respect to compute and cognitive labor, dividing them, taking the logarithm, and rearranging the terms, we arrive at the following equation:
\[
\ln\left(\frac{K_{it,\text{res}}}{H_{it}}\right) = \sigma \ln\left(\frac{w_{it}}{r_{it}}\right) + \sigma \ln\left(\frac{\gamma}{1-\gamma}\right).
\]

Therefore, we can estimate $\sigma$ by regressing $\ln(K_{it,\text{res}}/H_{it})$ on a constant and $\ln(w_{it}/r_{it})$ and looking at the coefficient on $\ln(w_{it}/r_{it})$. Intuitively, we can estimate how substitutable compute and labor are by seeing how the ratio of compute to labor changes as the relative price of labor to compute changes.

\subsection{Alternative CES Formulation in Frontier Experiments}

One potential problem with the baseline CES production function is that the required ratio of compute to labor in research does not depend on the frontier model size. Intuitively, as frontier models get larger, the compute demands of AI research should get larger as the firm needs to run near-frontier experiments. To accommodate this intuition, we will explore a re-parametrization of CES as an extension to our main, baseline results.

Let $E_{it} = x \, \frac{K_{it,\text{res}}}{K_{it,\text{tra}}}$ denote the number of near-frontier experiments a firm can run at time $t$. $\frac{K_{it,\text{res}}}{K_{it,\text{tra}}}$ is literally the number of frontier research training runs possible\footnote{Note that algorithmic advances do not improve this ratio because they improve effective training compute and effective research compute proportionally.}. $x \ge 1$ denotes the productivity benefit of extrapolating results from smaller experiments. For example, if you can accurately extrapolate experiments at $\tfrac{1}{1000}$ of frontier compute then $x = 1000$\footnote{To spell this out further, if research compute and training compute are equal, then by default you can run one experiment at the frontier. However, if you can extrapolate from experiments $\tfrac{1}{1000}$ the size, then you can run 1000 experiments, so $x = 1000$. Further, the value of $x$ does not really matter. If $x$ is a fixed number (e.g., AIs are not better at extrapolating than humans), then $x$ does not change the conditions under which there is an intelligence explosion and it does not change the estimate of $\sigma$.}.

Now the change in algorithm quality over time is given by
\[
\dot{A}_{it} = \theta A_{it}^{\phi} F(E_{it}, H_{it})^{\lambda}.
\]

We continue to suppose $F(\cdot)$ is CES. Following the same derivation steps as before, we get the following modified estimation equation:
\[
\ln\left(\frac{K_{it,\text{res}}}{H_{it}}\right) = \big[\sigma \ln\tfrac{\gamma}{1-\gamma} - (1-\sigma)\ln x\big] + \sigma \ln\left(\frac{w_{it}}{r_{it}}\right) + (1-\sigma)\ln K_{it,\text{tra}}.
\]

We can estimate this equation by regressing $\ln(K_{it,\text{res}}/H_{it})$ on a constant, $\ln(w_{it}/r_{it})$, and $\ln K_{it,\text{tra}}$. We will take the coefficient on $\ln(w_{it}/r_{it})$ as our estimate for $\sigma$.

\section{Data}\label{data}
Our data covers OpenAI from 2016--2024, Anthropic from 2022--2024, DeepMind from 2014--2024 and DeepSeek from 2023--2024. All prices are inflation-adjusted to 2023 USD. We use the following data sources.

$H_{it}$: We use headcount estimates from PitchBook, which provides data at a high frequency (roughly once per year). Unfortunately, the data does not make a distinction between research/engineering staff and operations/product staff. Assuming the ratio of research to operations staff remains constant over time, our results will be unbiased.

$K_{it,\text{tra}}$: We first take estimates of $K_{it,\text{tra}}$ from Epoch's notable models page, aggregating to the firm-year level by summing all training compute used across models in a given year. In cases where firms do not release (major) models in a given year, we assume training compute is the same as the prior year.

$K_{it,\text{res}}$: The Information reported that the ratio of OpenAI's research to training compute spend was $1:3$ in 2024. Therefore, we multiply our estimate of $K_{it,\text{tra}}$ by $\tfrac{1}{3}$ to get our estimate of $K_{it,\text{res}}$. This is a significant limitation as we are assuming that the fraction of research to training compute is a constant $\tfrac{1}{3}$ across firms and times\footnote{If this ratio has changed a lot over time, both of our estimates could be wrong. But in particular, the frontier experiments estimate could be badly wrong because we are essentially assuming that one of the inputs, number of frontier experiments, has been constant over time.}. This is the coarsest of all our variables.

$w_{it}$: Our most reliable wage data comes from DeepMind and OpenAI's financial statements which include total spend on staff. Combined with our estimates of headcounts, we can recover average wages. DeepMind's financial statements cover 2014--2023, while OpenAI's statements cover its period as a nonprofit from 2016--2018. We fill in the rest of the years and firms using data from firms' job postings, Glassdoor, H1B Grader, levels.fyi, news sources, and BOSS Zhipin. We specifically look for and impute the wage of level III employees (scientific researchers) at each firm. We use salary instead of total compensation and assume that salary constitutes 40\% of total compensation.\footnote{As long as the ratio of total compensation to salary has been constant over time, it does not matter which we use.}. While the data from financial statements is reliable, the imputed data involves a significant degree of estimation.

$r_{it}$: We use the rental rate of GPUs according to Epoch's data. We match each AI firm with its cloud provider (e.g., OpenAI with Microsoft Azure) and use the corresponding rental rate. In reality, AI firms buy many GPUs, although in a competitive market the depreciated, present-discounted price should match the hourly rental rate\footnote{Note, however, the AI GPU market is not competitive, as Nvidia owns a huge fraction of the market. However, this Epoch paper finds that pricing calculation via ownership vs. rental rates are fairly similar.}. We adjust for GPU quality by measuring the price in units of total FLOPs (e.g., FLOP/s times 3600 seconds in an hour) per dollar. We match firm-years to the GPUs that they were likely using at the time (e.g., OpenAI using A100s in 2022 and H100s in 2024). There is guesswork involved in the exact mix of GPUs that each firm is using in each year.

\begin{figure}[htbp]\centering
\includegraphics[width=\linewidth]{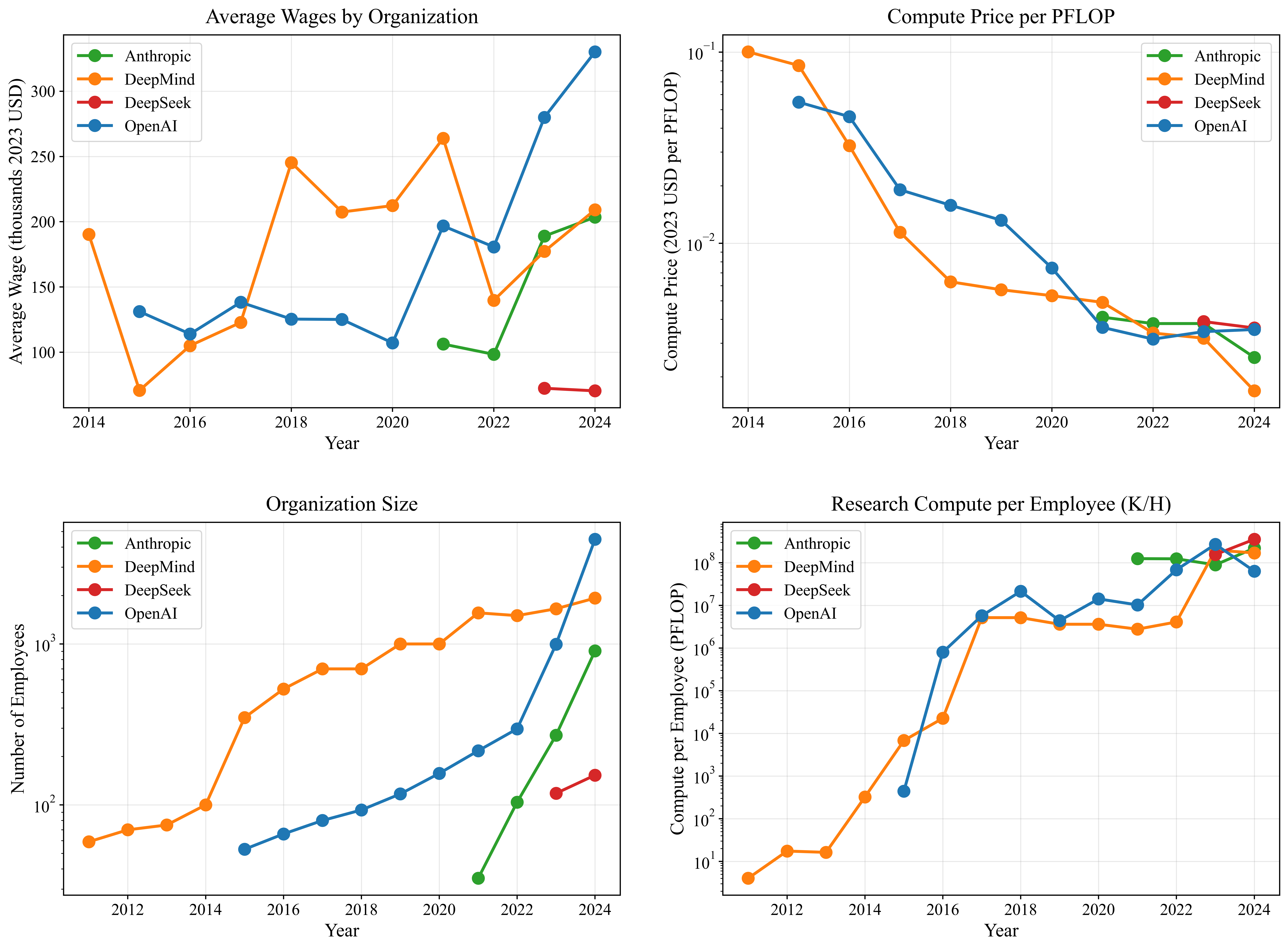}
\caption{Time trends}
\label{fig:1}
\footnotesize \textbf{Note:} This figure shows trends from 2012 to 2024 in four key variables for frontier AI firms: Anthropic, DeepMind, DeepSeek, and OpenAI. All monetary values are adjusted to 2023 USD.
\end{figure}

Figure \ref{fig:1} shows the time trend of average wage, compute price, the number of employees, and the research compute per employee by organization.

\section{Estimation Results}\label{results}

We estimate two sets of main results, one for the CES in compute specification and one for CES in frontier experiments specification. For both sets of results, we include firm fixed effects to correct for any time-invariant productivity differences between firms.

\begin{table}[htbp]\centering
\caption{Elasticity of Substitution Estimates}
\label{tab:elasticity}
\begin{tabular}{lcc}
\toprule
 & (1) & (2) \\
 & CES in Compute & CES in Frontier Experiments \\
\midrule
Elasticity of Substitution ($\sigma$)       & 2.583   & $-0.103$ \\
                       & (0.341) & (0.176) \\
                       & [0.657] & [0.419]   \\
\midrule
Controls                                    & Firm FE & Firm FE + $\ln(K_{\text{train}})$ \\
Observations                                & 27      & 27      \\
$R^2$                                       & 0.857   & 0.982   \\
\bottomrule
\end{tabular}
\vspace{0.5em}
\begin{minipage}{\linewidth}
\footnotesize
Notes: Observations are at the firm-year level. Column (1) uses a CES-in-compute specification. Column (2) includes $\ln(K_{\text{train}})$ to account for frontier experiment scale. All regressions include firm fixed effects. Standard errors in the paratheses are clustered at the firm level. Gamma estimates are constrained in closed form from first-order conditions. Monte Carlo standard errors are in the square brackets. 
\end{minipage}
\end{table}

 Table \ref{tab:elasticity} shows that the choice of model specification significantly affects the results. In the CES in compute specification, we estimate $\sigma = 2.58$, which implies research compute and cognitive labor are highly substitutable. However, in the frontier experiments case, we estimate $\sigma = -.10$. It is impossible in the economic model to have $\sigma < 0$, although this estimate is statistically indistinguishable from zero. Therefore, this result implies that frontier experiments and labor are highly complementary. Recall $\sigma = 0$ would denote perfect complements, where increasing cognitive labor without a corresponding increase in near-frontier experiments would result in zero growth in algorithm quality.

\begin{figure}[htbp]\centering
\includegraphics[width=\linewidth]{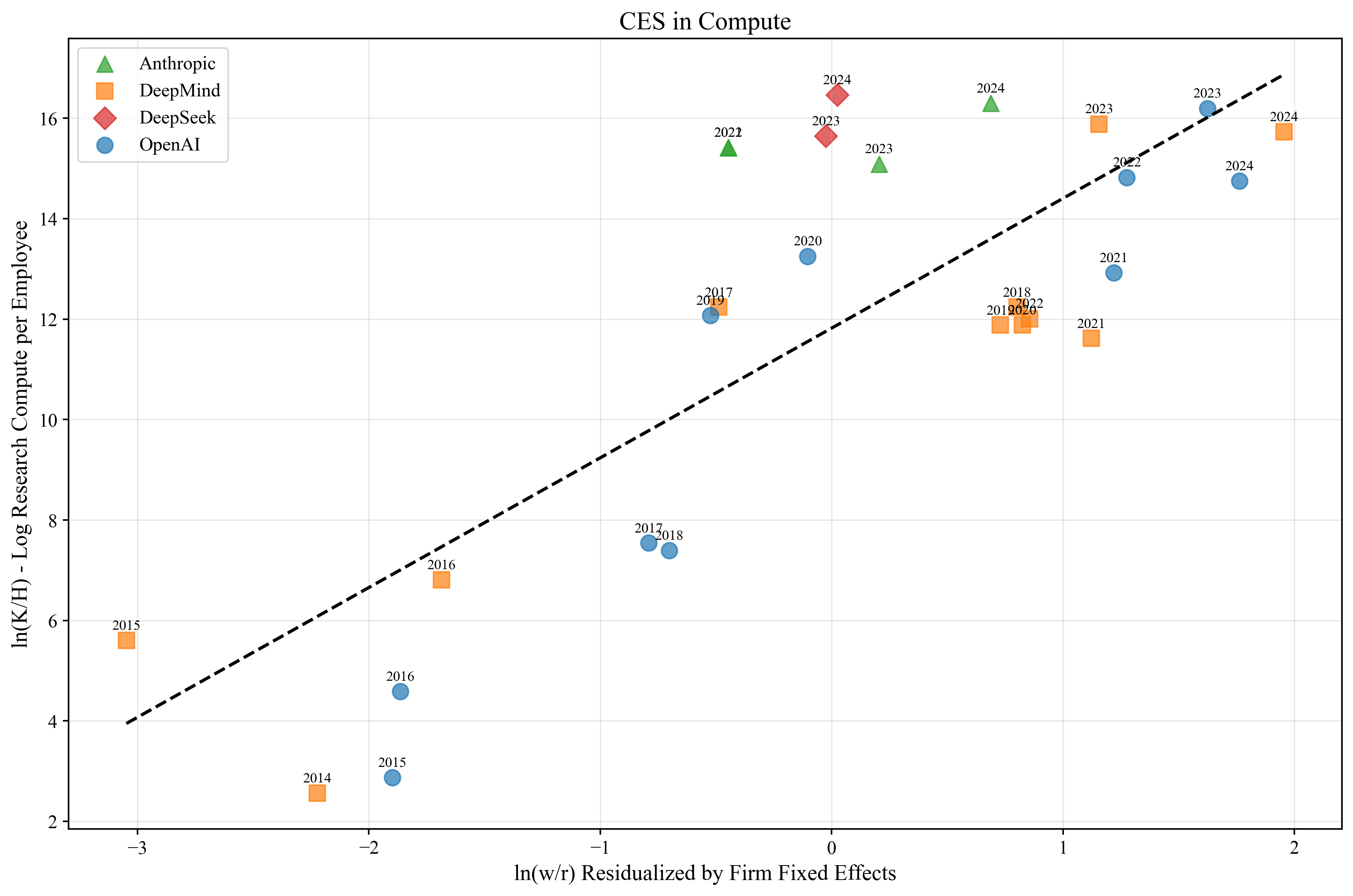}
\caption{Added-Variable Plot for the Baseline Model}
\label{fig:2}
\end{figure}

To get a visual understanding of these fits, figure \ref{fig:2} plots the regression result for the CES in compute specification. The slope of the fitted line corresponds to the estimated $\sigma$.

\begin{figure}[htbp]\centering
\includegraphics[width=\linewidth]{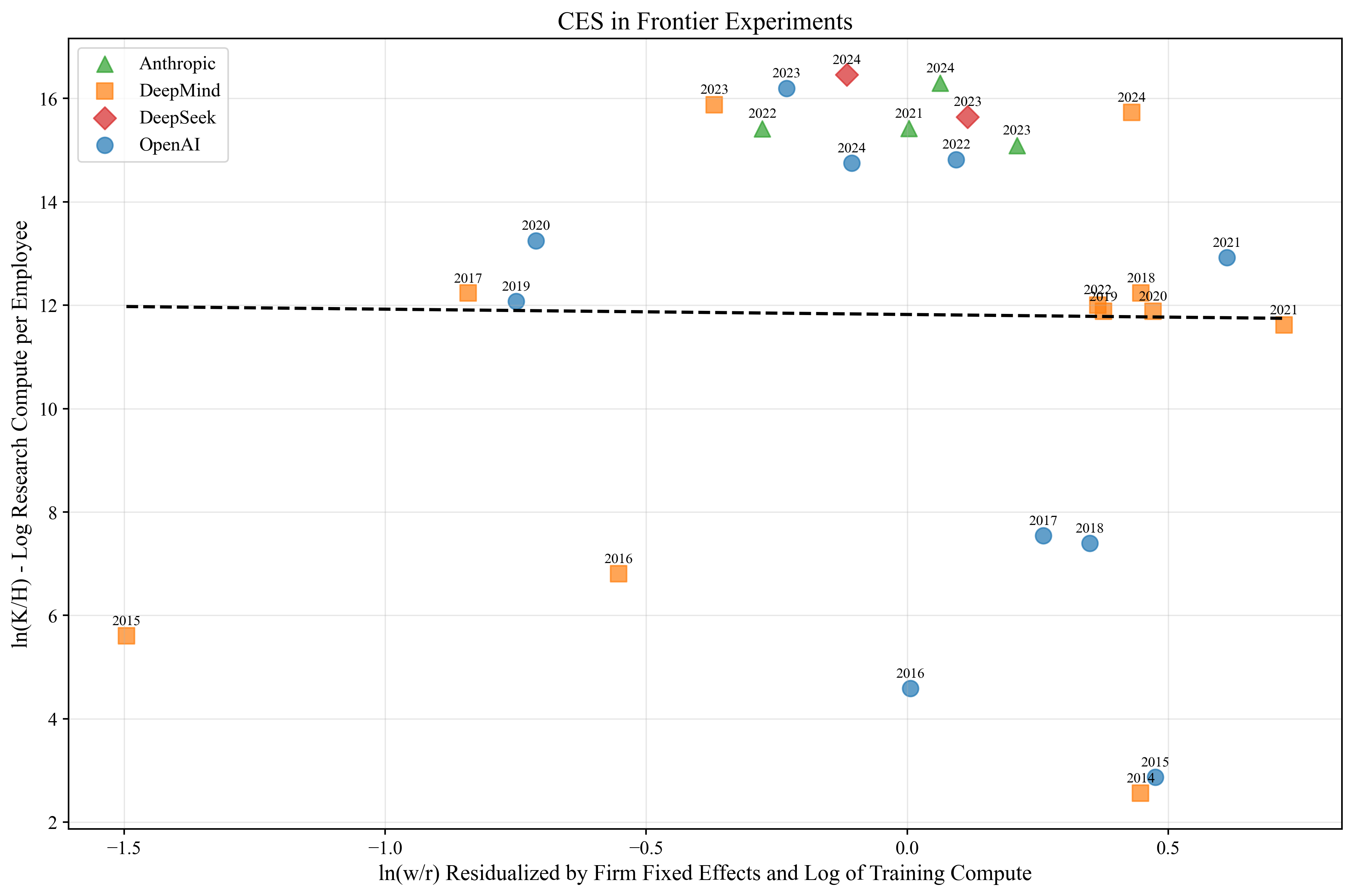}
\caption{Added-Variable Plot for the Frontier Experiments Model}
\label{fig:3}
\end{figure}

Figure \ref{fig:3} corresponds to the regression result for CES in frontier experiments.

We also performed some basic robustness tests in Section \ref{robust}. We get qualitatively similar results in all cases that we tried (substitutes in the CES in compute specification, complements in the CES in frontier experiments specification). 

Intuitively, the two models are expected to yield different results. Figure \ref{fig:1} shows that compute price decreases dramatically compare to the wage. A decrease in wage share means compute and labor are substitutes in the baseline model. But if instead of compute, we consider the number of frontier experiments, then the price for each experiment increases faster than wage. That means a decrease in wage share means frontier experiment and labor are complements.

Each specification has its advantages and disadvantages in terms of which specification to believe in. On the one hand, the baseline CES model has relatively better data. On the other hand, if the raw compute version was properly specified, then adding the training compute as a control should not change the coefficient on $\frac{w_{it}}{r_{it}}$.

We interpret these results with caution, as this analysis has several potential limitations. It is not obvious which specification is correct, the underlying data has reliability problems, and the data is from only four firms across a limited number of years. On a more technical level, a large amount of variation is explained by firms scaling up training compute over time, there is endogeneity/simultaneity bias\footnote{We have a version of this analysis where we use local wages as an instrument to address potential endogeneity. We get similar results to the OLS version, but we omit them for brevity and because we are still thinking about better instruments.}, and our analysis relies on simplifying assumptions such as the CES functional form and homogeneous, non-quality differentiated labor.

\section{conclusion} \label{conclusion}
This paper develops a theoretical framework to clarify the necessary conditions under which recursive self-improvement leads to explosive growth in algorithmic quality. We show that the elasticity of substitution between research compute and cognitive labor, $\sigma$, is a central determinant of whether such an outcome is plausible.

Empirically, we estimate $\sigma$ using a novel panel dataset of four leading AI firms from 2014 to 2024. Our baseline model suggests that compute and labor are strong substitutes, implying that recursive self-improvement could plausibly accelerate without being bottlenecked by compute. However, in our alternative model—which accounts for the increasing demands of frontier-scale experiments—we estimate that compute and labor are strong complements ($\sigma \approx 0$), indicating that gains in cognitive labor alone may not suffice to drive explosive progress without proportional increases in compute capacity. 

These results imply that the feasibility of a software-only intelligence explosion is highly sensitive to the structure of the AI research production function. If progress hinges on frontier-scale experiments, then compute constraints may remain a binding bottleneck, even as AI systems take over cognitive labor. Conversely, if smaller-scale experiments can be effectively leveraged or extrapolated, recursive self-improvement may proceed with less dependence on compute.

We caution that our findings should be interpreted with care. The analysis is limited by data availability, particularly in the construction of research compute and wage measures. Moreover, the small sample size and modeling assumptions (e.g., CES functional form, constant factor shares) impose constraints on inference. Nonetheless, this paper provides a first empirical estimate of the substitutability between compute and labor in frontier AI R\&D and lays a foundation for future work on this question. In future studies, we plan to dive deeper into understanding AI research's production function and conduct RCTs to recover the elasticity of substitution. We also plan to estimate other parameters including $\phi$ (whether ideas get harder to find) and $\lambda$ (parallelization penalty). 

\appendix
\renewcommand{\thetable}{A\arabic{table}}

\section{Robustness Checks for Elasticity Estimates} \label{robust}

We also performed some basic robustness tests, like restricting the sample to after 2020 or 2022 (after GPT-3 or GPT-4 released), excluding 2024 (for the concern that AI had started to meaningfully assist in AI research), restricting the sample to DeepMind only (where we have the highest quality wage data), and changing how we calculate wages or research compute. See Tables \ref{tab:2020}, \ref{tab:excl2024}, \ref{tab:deepmind}, and \ref{tab:adjustedcompute}. 

\begin{table}[htbp]\centering
\caption{Robustness: Sample Restricted to 2020 Onwards}
\label{tab:2020}
\begin{tabular}{lcc}
\toprule
 & (1) & (2) \\
 & CES in Compute & CES in Frontier Experiments \\
\midrule
Elasticity of Substitution ($\sigma$)       & 1.486   & $-0.769$ \\
                       & (0.672) & (0.464) \\
                       & [0.633] & [0.567]   \\
\midrule
Controls                                    & Firm FE & Firm FE + $\ln(K_{\text{train}})$ \\
Observations                                & 16      & 16      \\
$R^2$                                       & 0.562   & 0.886   \\
\bottomrule
\end{tabular}
\vspace{0.5em}
\begin{minipage}{\linewidth}
\footnotesize
Notes: Observations are at the firm-year level. Column (1) uses a CES-in-compute specification. Column (2) includes $\ln(K_{\text{train}})$. All regressions include firm fixed effects. Standard errors in parentheses are clustered at the firm level. Monte Carlo standard errors are in square brackets.
\end{minipage}
\end{table}

\begin{table}[htbp]\centering
\caption{Robustness: Excluding Year 2024}
\label{tab:excl2024}
\begin{tabular}{lcc}
\toprule
 & (1) & (2) \\
 & CES in Compute & CES in Frontier Experiments \\
\midrule
Elasticity of Substitution ($\sigma$)       & 2.638   & $-0.129$ \\
                       & (0.449) & (0.128) \\
                       & [0.449] & [0.418]   \\
\midrule
Controls                                    & Firm FE & Firm FE + $\ln(K_{\text{train}})$ \\
Observations                                & 23      & 23      \\
$R^2$                                       & 0.835   & 0.992   \\
\bottomrule
\end{tabular}
\vspace{0.5em}
\begin{minipage}{\linewidth}
\footnotesize
Notes: Same specification as main results, but excludes 2024 observations. See notes to Table~\ref{tab:2020}.
\end{minipage}
\end{table}

\begin{table}[htbp]\centering
\caption{Robustness: DeepMind Subsample Only}
\label{tab:deepmind}
\begin{tabular}{lcc}
\toprule
 & (1) & (2) \\
 & CES in Compute & CES in Frontier Experiments \\
\midrule
Elasticity of Substitution ($\sigma$)       & 2.297   & $-0.007$ \\
                       & (0.556) & (0.281) \\
                       & [0.810] & [0.422]   \\
\midrule
Controls                                    & None & $\ln(K_{\text{train}})$ \\
Observations                                & 11      & 11      \\
$R^2$                                       & 0.820   & 0.995   \\
\bottomrule
\end{tabular}
\vspace{0.5em}
\begin{minipage}{\linewidth}
\footnotesize
Notes: Subset of data using only DeepMind observations, where wage data is highest quality. Column (1) has no controls; Column (2) includes $\ln(K_{\text{train}})$.
\end{minipage}
\end{table}

\begin{table}[htbp]\centering
\caption{Robustness: Adjusted Compute Cost Specification}
\label{tab:adjustedcompute}
\begin{tabular}{lcc}
\toprule
 & (1) & (2) \\
 & CES in Compute & CES in Frontier Experiments \\
\midrule
Elasticity of Substitution ($\sigma$)       & 0.893   & $-0.127$ \\
                       & (0.103) & (0.116) \\
                       & [0.123] & [0.155]   \\
\midrule
Controls                                    & Firm FE & Firm FE + $\ln(K_{\text{train}})$ \\
Observations                                & 27      & 27      \\
$R^2$                                       & 0.879   & 0.983   \\
\bottomrule
\end{tabular}
\vspace{0.5em}
\begin{minipage}{\linewidth}
\footnotesize
Notes: This table modifies the specification by adjusting how compute prices are calculated. Estimates remain qualitatively similar.
\end{minipage}
\end{table}

\section{Instrumental Variable}
An unobserved confounding variable, such as advancements in general labor productivity within AI research, could concurrently increase the relative price of specialized labor (if such labor becomes more valuable and scarce) and also drive up the utilization of compute per employee as researchers become more efficient at leveraging computational resources. In such a scenario, the observed positive correlation might not solely reflect substitution due to relative price changes. 

To mitigate potential endogeneity bias, we adopt an instrumental variable (IV) strategy. We instrument for the endogenous wage-to-compute-cost ratio ($w_{it}/r_{it}$) with the local, exchange-rate-adjusted wage level for skilled labor. The validity of this instrument hinges on two standard assumptions. First, the instrument must be relevant, meaning it is correlated with the endogenous variable. This condition is met as the prevailing local wage is a strong predictor of the wages AI firms must pay to attract researchers. Second, the instrument must satisfy the exclusion restriction, meaning it only affects the outcome variable—the research compute-to-labor ratio ($K_{it,\text{res}}/H_{it}$)—through its effect on the wage-to-compute-cost ratio. We argue this holds because broader wage trends in a specific geographic location are unlikely to directly influence the technological production function of a global AI firm in a short period of time, other than by altering the relative cost of labor.

Table \ref{tab:iv_results} shows that the results do not change much with the IV approach. 
\begin{table}[htbp]\centering
\caption{Instrumental Variable Estimates for Elasticity of Substitution}
\label{tab:iv_results}
\begin{tabular}{lcc}
\toprule
 & (1) & (2) \\
 & CES in Compute & CES in Frontier Experiments \\
\midrule
Elasticity of Substitution ($\sigma$) & 2.768   & 0.126   \\
                                      & (0.309) & (0.384) \\
\midrule
Controls                              & Firm FE & Firm FE + $\ln(K_{\text{train}})$ \\
Observations                          & 27      & 27      \\
$R^2$                                 & 0.853   & 0.981   \\
First-Stage F-statistic               & 124.94  & 9.88    \\
\bottomrule
\end{tabular}
\vspace{0.5em}
\begin{minipage}{\linewidth}
\footnotesize
Notes: Observations are at the firm-year level. The endogenous wage-to-compute-cost ratio is instrumented by local, exchange-rate-adjusted wage levels. Column (1) is the CES-in-compute specification; Column (2) is the CES-in-frontier-experiments specification. All regressions include firm fixed effects. Standard errors, clustered at the firm level, are in parentheses.
\end{minipage}
\end{table}

\bibliographystyle{econometrica}

\bibliography{reference.bib}

\end{document}